\documentstyle[twocolumn,aps,epsfig,floats]{revtex}

\def\etal{{\em et al.}}

\begin{document}
\bibliographystyle{abbrv}



\title{Protein structure prediction as a hard optimization problem:\\
the genetic algorithm approach}

\author{Mehul~M.~Khimasia \footnote{mmlk2@cam.ac.uk}\\}
\address{Theory of Condensed Matter,
Cavendish Laboratory,\\
University of Cambridge,
Madingley Road,
Cambridge CB3 0HE, U.K.\\
telephone: +44-1223-337004, facsimile: +44-1223-337356\\}

\author{Peter~V.~Coveney \footnote{coveney@cambridge.scr.slb.com} \\}
\address{Schlumberger Cambridge Research,\\
High Cross,
Madingley Road,
Cambridge CB3 0EL, U.K.\\}

\vspace{0.5cm}

\date{\today}

\maketitle

\begin{abstract}
  Protein structure prediction can be shown to be an NP-hard
  problem; the number of conformations grows exponentially with the
  number of residues. The native conformations of proteins occupy a
  very small subset of these, hence an exploratory, robust search
  algorithm, such as a genetic algorithm (GA), is required. The
  dynamics of GAs tend to be complicated and problem-specific.
  However, their empirical success warrants their further study.  In
  this paper, guidelines for the design of genetic algorithms for
  protein structure prediction are determined.  To accomplish this,
  the performance of the {\em simplest} genetic algorithm is
  investigated for simple lattice-based protein structure prediction
  models (which is extendible to real-space), using energy
  minimization.  The study has led us to two important conclusions for
  `protein-structure-prediction-genetic-algorithms'.  Firstly, they
  require high resolution building blocks attainable by multi-point
  crossovers and secondly they require a local dynamics operator to
  `fine tune' good conformations. Furthermore, we introduce a
  statistical mechanical approach to analyse the genetic algorithm
  dynamics and suggest a convergence criterion using a quantity
  analogous to the free energy of a population.
\end{abstract}

\vspace{1.0cm}

\noindent
{\bf{Short title: Genetic algorithm approach to structure prediction}}

\vspace{0.5cm}


\vspace{0.5cm}

\noindent
{\bf{Abbreviations: PSP protein structure prediction; GA genetic
algorithm; PSP-GA protein-structure-prediction-genetic-algorithm; MC
Monte Carlo; GA-MC genetic algorithm-Monte Carlo; HZ hydrophobic
zipper; REM Random Energy Model}}

\newpage


\section{Introduction}

Conceptually, proteins fold from their 1D polymer chain of amino acids
(primary structure) to 3D stable, `unique' conformations (tertiary
structure).  Anfinsen \cite{Anf73} showed that folding requires
knowledge of the amino acid sequence alone; the determination of the
native (biologically functioning) structure from its sequence is known
as the {\em protein folding} problem.  Much research has gone into
elucidating the folding dynamics \cite{KarSha92:ProtFold} but a
practical theory is still beyond our understanding. 

The folding problem attempts to understand the dynamics of the
folding - {\em how} the sequence of amino acids arrive at the native
state. It is clear that a full solution to this problem would be able
to predict the tertiary structure of an amino acid sequence.
Unfortunately, such a solution is beyond our understanding at
present. 

However, amino acid sequences are effectively being determined at a
higher rate than that of their corresponding native structures.  Since
knowledge of the native structure is important to understanding the
function of a protein, a potentially more practical problem is
protein structure prediction.  There is a possible confusion in
terminology between the two terms, {\em protein folding} and {\em
  protein structure prediction},as researchers use the terms
loosely and interchangeably.
The difference must be stressed as there are `folding
simulations', in the literature, which carry out a conformational
search in a manner that is not attributable to the true folding
dynamics of proteins (e.g.  \cite{UngMou93:2d}).

Protein folding mainly concerns the dynamics of the problem which
experimentalists study using hydrogen-exchange techniques, for
example; theorists often use computer simulations of simple models,
often on lattices, to elucidate the folding dynamics.  Protein
structure prediction, however, is only interested in the end result;
experimentalists often use crystallisation techniques coupled with
x-ray diffraction to determine the tertiary structure, whereas
theorists tend to use computer-based optimization methods.  This is
the approach discussed in this paper.  The problem involves two
aspects: (1) the specification of the function to optimize and (2) the
choice of a search algorithm.

The prediction of protein structures using optimization methods have
been more successful using comparative modelling \cite{MoultEtal95}
techniques which include sequence alignment, threading \cite{Tay88}
and the use of secondary structural propensities \cite{GaOsRo78}.
However, a physical approach based on energy minimization is used
here.  It is clear that the interactions between the residues and the
solvent molecules drive the protein towards its native state.
Determining the fundamental interactions is important, not only for
protein structure prediction, but also for the protein folding
problem.  A further reason for the physical approach is that
comparative modelling requires the known structure of one or more
homologous proteins - these cannot always be guaranteed to exist.

The second aspect of the protein structure prediction problem is
concerned with designing a search strategy for the energy
minimization. It is clear that a blind search through the
conformational space is impossible, as it would take a time greater
than the age of the universe.  A similar `paradox' for real proteins,
based on observed protein folding times ($\sim 10^{-3}s$), was first
noted by Levinthal \cite{Lev68}.  Furthermore, several authors have
proved various models of protein structure prediction to be
computationally NP-hard \footnote{A good introduction to the theory of
  NP-completeness and NP-hardness can be found in \cite{GarJoh79}.}
\cite{UngMou93:NP,Fra93,NgoMar92}. This implies that no efficient
algorithm can be designed to guarantee finding the native state
amongst the exponentially many. A related idea to the complexity of
the conformational search is the concept of `rugged' energy landscapes
\cite{BrOnSoWo95}.  Although optimal solutions are not guaranteed,
robust, exploratory, non-deterministic search algorithms (e.g.
simulated annealing \cite{KiGeVe83} and genetic algorithms
\cite{Hol75}) can locate good, near-optimal solutions, within a
reasonable time. In the present paper, we focus on the design of a
good conformational search strategy for the problem, leaving the
discussion of the best choice of energy functions for protein
structure prediction to a later paper.

Genetic algorithms were invented by John Holland \cite{Hol75} in his
quest for a theory of adaptive processes.  The concept was inspired by
Darwin's evolutionary theory (loosely `survival of the fittest') and
in particular `neo-Darwinism' according to which genetic
recombinations and mutations play a dominant role in the evolution of
a species.  It is generally believed that Nature evolves so that
individuals that are the best adapted to their co-evolving environment
survive, while the poor ones die off; this is an example of
optimization, more commonly referred to as `adaptation' in biology
\cite{CovHig95:frontiers}.  Due to their highly nonlinear nature,
genetic algorithms are difficult to analyse.  There is no asymptotic
global optimum convergence proof, as there is in simulated annealing
\cite{AarKor89}, nor are there any general rules to design a GA for a
specific problem.  However, their empirical success for the solution
of numerous NP problems (\cite{Dav91:handbook} and references therein)
warrants their further study.

Previous applications of genetic algorithms to the protein structure
prediction problem
\cite{DanArg92,DanArg94,DanArg96,UngMou93:2d,UngMou93:3d,S-K92,SuThDi95}
have not considered the GA design issue.  As in any problem, the
simpler the algorithm is, the fewer the parameters it requires, the
easier it is to understand and improve the performance.  For this
purpose we use a modified version of Goldberg's \cite{Gol89:gasoml}
Simple Genetic Algorithm (SGA), written in C.  Dandekar and Argos
\cite{DanArg92,DanArg94,DanArg96} also use a version of the SGA but
their work differs to ours in that our objectives are different.
Firstly, they optimize the structural features of proteins (such as
$\alpha$-helices and $\beta$-strands) rather than adopting an energy
minimization approach.  Secondly, little comparison was made with the
performance of other algorithms.  Schulze-Kremer \cite{S-K92} wrote an
elaborate genetic algorithm to optimize real-space dihedral angles for
a fully atomistic representation of proteins, based on CHARMM
\cite{BroEtal83} energy minimization. Unfortunately, results were
poor.  Clearly, our principal research aim is to predict realistic
structures as well, but only when a good method has been established.
Unger and Moult \cite{UngMou93:2d,UngMou93:3d} compared a Monte Carlo
search with a `genetic algorithm', using simple exact lattice models.
However, their genetic algorithm is, strictly speaking, a hybrid GA. It
incorporates several Monte Carlo conformers with the occasional
crossover between structures.  For this reason, we call their approach
a `Genetic Algorithm-Monte Carlo' (GAMC). They compared this method
with a Monte Carlo search and concluded that the GAMC found lower
energy solutions.

In this paper, we compare the Simple Genetic Algorithm with work by
Unger and Moult \cite{UngMou93:3d}; Yue \etal \cite{Yueetal95} and
Sali \etal \cite{ShFaGuKa91} and determine guidelines for designing
protein-structure-prediction-genetic-algorithms.  

The paper is structured in the following way.  Section \ref{WhyGA}
highlights the conformational search issue and the need for a genetic
algorithm approach.  Following that, section \ref{DesignGA} provides a
description of the method for determining guidelines for GA design.
Section \ref{SGA} discusses the Simple Genetic Algorithm used for the
lattice conformational search.  The HP-model \cite{Dilletal95} and
REM-model \cite{ShaGut89} are described in section
\ref{EnergyFunctions}, as are the various methods used to minimize
these potentials \cite{Yueetal95,UngMou93:2d,UngMou93:3d,ShFaGuKa91}.
Lattice conformational search results from the SGA are compared with
the other methods in section \ref{Results}, while section
\ref{Discussion} provides a discussion of the `protein structure
prediction-genetic algorithm' (PSP-GA) design principles that have
emerged from this work.

\section{Why Genetic Algorithms for Protein Structure Prediction?}
\label{WhyGA}

Protein structure prediction is analytically difficult to solve. The
problem is thought to stem from the exponential nature of the
conformational search space.  The number of conformations of a protein
with $N$ amino acid residues grows exponentially as ${\gamma}^N$ where
$\gamma$ is the average number of conformations per residue (typically
$\sim$10).  This suggests that an algorithm would require an
exponential time to search the whole conformational space for the
native state.  

However, problems with exponentially growing search spaces are not new
in physics (e.g. ideal gases) but many are solvable due to the
symmetries and conservation laws that can be exploited.  With proteins
there is an added difficulty that the interactions are complex - it is
not clear whether there are enough symmetries to reduce the problem to
a tractable solution.  In addition to this, proteins, although
macromolecular, do not contain $\sim 10^{23}$ atoms to guarantee a
valid statistical mechanical and thermodynamic treatment.

Furthermore, work by Unger and Moult \cite{UngMou93:NP}; Fraenkel
\cite{Fra93}; Ngo and Marks \cite{NgoMar92} suggests that the protein
structure prediction problem is NP-hard, that is computationally
impossible to guarantee an exact solution.  Bryngelson, Wolynes \etal
\cite{BrOnSoWo95}, for example, advocate that the energy landscape of
proteins must be `rugged'.  This reflects the various energy barriers
that have to be crossed and thus the hurdles that a conformational
search algorithm must be able to deal with.  

For these reasons `intelligent' conformational search algorithms have
become popular in structure prediction.  Unlike gradient-based
methods\cite{MacCroHag90}, which tend to terminate at local minima, genetic
algorithms `hop' around the conformational space independent of local
derivatives.  A selection process focuses the search in low energy
areas, whereas a recombination stage maintains exploration of the
search space.

\section{Designing Genetic Algorithms for Protein Structure Prediction}
\label{DesignGA}
A genetic algorithm is made up of 4 basic components: representation;
selection; recombination and evaluation.  Representation deals with
formulating the specific problem as a digital string of parameters.
This, combined with the evaluation function (in our case
conformational energy) describes the optimization problem.  The
remaining two components, selection and recombination, provide the
dynamics of the GA search, which drive the population of solutions
towards the global optimum. Within each unit, there are several
options leading to numerous variations of genetic algorithms; some
examples and corresponding parameters are listed in table I.  To
determine guidelines for designing PSP-GAs, we used the following
principles:

\begin{enumerate}
\item A first approach should always be the simplest approach. In
  order to analyse the GA dynamics, we used a GA with the simplest
  options and with the fewest parameters - a Simple Genetic Algorithm
  - to search lattice conformational space.

\item A systematic sample search of the parameter space was carried out
  to determine optimal parameter values for the SGA.

\item Having calibrated the SGA in step 2, we compared the SGA
  conformational search ability using several test energy functions.

\item Conformations, and their corresponding energies, generated by the
  SGA were compared with conformations predicted by other search
  methods. 

\item The time evolution of conformations generated by the SGA were
  observed.  Minimal requirements and improvements for PSP-GAs are
  proposed.
\end{enumerate}

\begin{table}[!htb]
\begin{center}
\smallskip
\begin{tabular}{l}
  {\em GA Options and Parameters} \\ \hline Population: static or
  variable size \\ Representation of solutions: bit string, reals,
  symbolic \\ Maximum number of generations or convergence criteria \\ 
  Recombination operators: 1-pt crossover, uniform crossover,\\ 
  mutation, perturbation \\ Recombination probabilities: static,
  variable or dynamic \\ Selection methods: roulette, tournament,
  rank, elitism \\ Fitness scaling: linear with cut-off, quadratic,
  exponential
\end{tabular}
\caption{Examples of various options in designing a genetic algorithm.}
\label{tabparam}
\end{center}
\end{table}

The merit of simple exact lattice models \cite{Dilletal95} is the
ability to test ideas easily and suggest extrapolations to real
systems. Simple exact models are `simple' since only a few
parameters are required, and `exact' since physical properties can be
calculated exactly.  Lattice models, although unrealistic in
appearance, provide several advantages over real-space models.  From a
folding dynamics point of view, they can explore long time behaviour;
while from a structure prediction/optimization point of view, as in
our case, they provide a valuable test-bed for protein structure
optimizers.  

Runs with various initial conditions were carried out to ensure that
the GA produced similar results each time.

\section{The Simple Genetic Algorithm for Lattice Conformational Search}
\label{SGA}
The Simple GA (SGA), as defined by Goldberg \cite{Gol89:gasoml}, is
the simplest of all genetic algorithms.  The original SGA manipulated
binary strings which encode a trial solution of the problem at hand.
However, a major modification for searching protein conformations on a
cubic lattice is to use a more natural representation for this
problem. Since a simple cubic lattice is spatially restricting, a
string of bond directions represents a folded chain of beads; each
symbol corresponds to an increment or decrement in the appropriate
Cartesian coordinate of the successive monomer beads (see table
\ref{latticerep}).  A conformation of the polymer is then translated
to a set of monomer positions ${\bf r}_i$ $(i=1,\cdots,N)$ (where $N$
is the number of monomer beads (residues)).

\begin{table}[!htb]
\begin{center}
\begin{tabular}{lc}
Direction & $\Delta${\bf r}\\ \hline
(U)P & $r_z\mapsto r_z+1$ \\
(L)EFT & $r_x\mapsto r_x-1$ \\
(F)RONT & $r_y\mapsto r_y+1$ \\
(B)ACK & $r_y\mapsto r_y-1$ \\
(R)IGHT & $r_x\mapsto r_x+1$ \\
(D)OWN & $r_z\mapsto r_z-1$ \\
\end{tabular}
\caption{Bond directions describing lattice conformations. A bond
  direction corresponds to a change,$\Delta{\bf r}$, in one of the
  Cartesian coordinates of the successive monomer, keeping all other
  coordinates the same as the previous monomer.}
\label{latticerep}
\end{center}
\end{table}

This representation has access to {\em all} $6^N$ lattice
conformations, including all the non-physical, non-self-avoiding
conformations.  Thus the search task is a formidable one.  A random
population of conformations is generated and manipulated according to
the GA dynamics (selection and recombination). The population size,
$\cal{S}$, is kept fixed. All individuals are replaced at each
iteration, except for two copies of the current best conformation -
this is known as `elitism'.

Selection is linearly proportional to fitness so that the probability,
$P_i$, of selecting the $i^{th}$ conformation, with a fitness value
$F_i$, to propagate to the next time step is given by:

\begin{equation}
P_i=\frac{F_i}{\sum_{j=1}^{\cal{S}}{F_j}}
\end{equation}

\noindent
Probabilities must be positive so a linear mapping with a cut-off value
is used to convert the energy ($E$) minimization problem to a fitness ($F$)
maximization:

\begin{equation}
 F_i = \cases{
 -E_i & if $E_i < 0$; \cr
   0  & if $E_i \geq 0$. \cr
 } 
\end{equation}

\begin{figure}[htbp]
\centerline{
\psfig{file=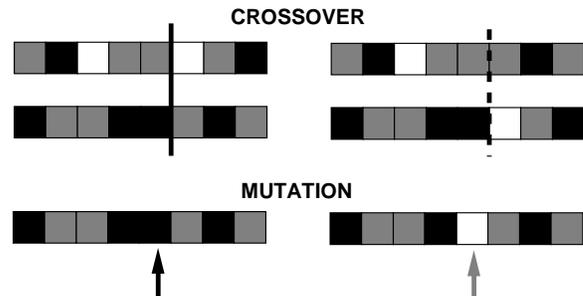, width=3in}
           }
\caption{The top half represents the action of a one-point crossover
operator. Below is an illustration of a single genewise mutation. In
both cases the left side represents the `before' situation, and
the right side, the `after' state.}
\label{oper}
\end{figure}

\begin{figure}[htbp]
\centerline{
\psfig{file=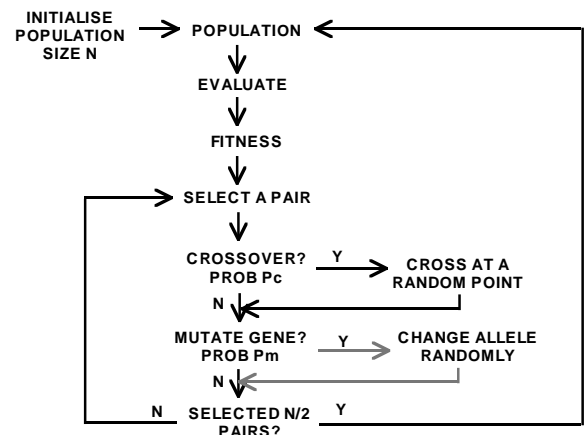,width=3in}
           }
\caption{The simple genetic algorithm.  A population of $N$
conformers is initialised.  These trial conformations are evaluated
and selected, based on their fitness.  Pairs of selected conformers are
crossed over with a probability, $P_c$.  Each bond direction (`gene') is
then randomly mutated with a probability, $P_m$.  A new population is
selected in this manner and repeated for many generations (typically
2000).}
\label{GAflow}
\end{figure}

Selected individuals, strings, are modified in a recombination
process, to generate new solutions.  Figure \ref{oper} shows a
schematic representation of the one-point crossover and gene-wise
mutation operators used by the SGA. These operators act stochastically
on the selected individuals with fixed probabilities.  In one-point
crossover, a random crossover point is chosen for a pair of selected
individuals, and the bond directions (`genes') are swapped up to the
crossover point.  Mutations act on a single individual and randomly
change the value of bond directions along the string.  The workings of
the SGA are summarised in figure \ref{GAflow}.

\section{Test Energy Functions}
\label{EnergyFunctions}
\subsection{Random Energy Model}

Originally used in spin glass theory, the Random Energy Model (REM)
\cite{Der80} was applied to protein folding by Bryngelson and Wolynes
\cite{BryWol87} and later formulated for lattice protein models by
Shakhnovich and Gutin \cite{ShaGut89}. In the REM, a protein is
described by a fixed sequence of random interaction energies. 
Although not discussed here, the random
interaction model lends itself for analytical studies of the
coil-globule transition in proteins \cite{ShaGut89}.
Specifically, the conformational energy is calculated as:

\begin{eqnarray}
E(\{{\bf r}_1..{\bf r}_N\})& = 
& \sum_{i,j=1; i<j}^N B_{ij} \Delta ({\bf r}_i-{\bf r}_j) + \nonumber \\
& & D_2\sum_{i,j}^N \delta({\bf r}_i-{\bf r}_j) + \nonumber \\
& & D_3\sum_{i,j,k}^N \delta({\bf r}_i-{\bf r}_j) \delta({\bf r}_j-{\bf r}_k)
\label{REM}
\end{eqnarray}
\\

\begin{center}
\begin{tabular}{l}
$\Delta ({\bf r}_i-{\bf r}_j) = \cases{ 
        1 & \quad \mbox{for i \& j nearest neighbours} \cr 
        0 & \quad \mbox{otherwise} \cr }$ 
\\
\\ 
$\delta ({\bf r}_i-{\bf r}_j) = \cases{ 
        1 & \quad \mbox{if monomers i,j occupy same site} \cr 
        0 & \quad \mbox{otherwise} \cr }$ 
\\

where,\\
$D_2    \equiv$ energetic penalty parameter for sites containing\\ 2 or more monomers \\
$D_3    \equiv$ energetic penalty parameter for sites containing\\ 3 or more monomers \\
$B_{ij} \equiv$ disorderly interaction energies \\
(energies are in units of $k_BT$ where $k_B$ is Boltzmann's \\ 
constant and $T$ is the temperature in Kelvin) \\
\end{tabular}
\end{center}

The interaction matrix, $B_{ij}$, is symmetric but randomly generated
with a Gaussian distribution:

\begin{equation}
P(B_{ij}) = (2 \pi B^2)^{- \frac{1}{2}} \exp(-(B_{ij}-B_0)^2/2B^2)
\end{equation}

The compactization observed in globule proteins is modelled in
Eq.(\ref{REM}) using $B_{ij}$ as a negative interaction potential with
mean $B_0$.  $B$ defines the spread, i.e. the standard deviation, of
the compactization interactions.  The greater the spread, the more
heterogeneous the protein.  A zero-spread corresponds to uniform
interactions; this special case constitutes the Fixed Energy Model
(FEM) which we have used for the larger polymers studied here (64mer,
125mer).

The last two terms in Eq.(\ref{REM}) represent excluded volume
effects, that is they are energetic {\em penalties} for conformations
with lattice sites occupied by more than two ($D_2$ term) or more than
three ($D_3$ term) monomers.  The objective of the genetic algorithm
is to find conformations without multiple occupancies at a single
site, by minimising this energy function.

\subsection{HP-model}

It is well known that correlations in the sequence of amino acid
residues lead to sub-structures (secondary structures) common to all
protein structures.  These correlations can reduce the size of the
`alphabet' (code) from 20 symbols to a lesser number.  The simplest
and most interesting is the classification of residues into two types: H
and P \cite{LauDil89}.  The energy function for this system favours
interactions between HH monomer types and is indifferent to all other
(PP and HP) interactions.  This is known as the HP-model. It aims to
highlight the importance of the hydrophobic effect in protein folding.
In an aqueous medium, globular proteins tend to have a core of
hydrophobic residues, surrounded by polar residues on the surface; in
the HP-model, the H monomers correspond to hydrophobic residues that
collapse to form a core surrounded by polar, P, monomers.  Much work
has been carried out by Dill and co-workers on this model
\cite{ChaDil91,YueDil93,SuThDi95,Yueetal95,Dilletal95}.

The energy function for this model is:

\begin{eqnarray}
E(\{{\bf r}_1..{\bf r}_N\})& = 
& -|\epsilon| \sum_{i,j=1; i<j}^N \Delta ({\bf r}_i-{\bf r}_j) + \nonumber \\
& & {\epsilon}_2\sum_{i,j}^N \delta({\bf r}_i-{\bf r}_j) + \nonumber \\
& & {\epsilon}_3\sum_{i,j,k}^N \delta({\bf r}_i-{\bf r}_j) \delta({\bf r}_j-{\bf r}_k)
\label{HP}
\end{eqnarray}

\begin{center}
\begin{tabular}{l}
$\Delta ({\bf r}_i-{\bf r}_j) = \cases{ 
        1 & \quad \mbox{if i,j both H-type \& nearest neigh.} \cr 
        0 & \quad \mbox{otherwise} \cr }$ 
\\
\\ 
$\delta ({\bf r}_i-{\bf r}_j) = \cases{ 
        1 & \quad \mbox{if monomers i,j occupy same site} \cr 
        0 & \quad \mbox{otherwise} \cr }$ 
\\

where,\\
$|\epsilon| \equiv$ strength of HH attraction (usually taken as 1)\\
$\epsilon_2    \equiv$ energetic penalty parameter for sites containing\\ two or more monomers \\
$\epsilon_3    \equiv$ energetic penalty parameter for sites containing\\ three or more monomers \\
\end{tabular}
\end{center}

Strictly speaking, the final two terms ($\epsilon_2, \epsilon_3$) in
Eq.(\ref{HP}) are not included in the HP-model. We borrow these terms
from the Random Energy Model to drive the conformational search
towards self-avoiding conformations.  This is necessary since {\em
  all} possible conformations are allowed; an energetic
penalty is required to penalise conformations with multiple
occupancies at a single site.  For both models (REM and HP), the
penalty terms reduce to zero if the search is successful and a low
energy self-avoiding conformation is found.

\section{Results}
\label{Results}
\begin{figure}[!htb]
\centerline{
\psfig{file=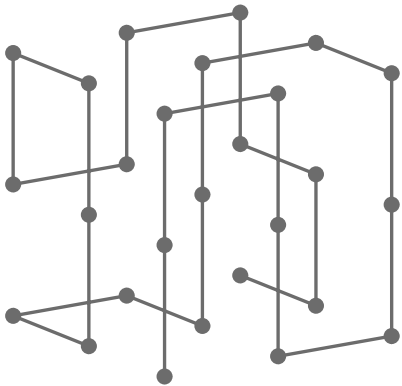, width=3in}
           }
\caption{Example of a 100\% cubic 27mer conformation found by the SGA.}
\label{27cubic}
\end{figure}

The first test function used was that from the Random Energy Model,
Eq.(\ref{REM}).  Shakhnovich \etal \cite{ShFaGuKa91} used a Metropolis
Monte Carlo algorithm with this energy function to find the native
states of 27mer heteropolymers. They interpreted the Monte Carlo
`folding' algorithm as modelling the folding dynamics of polymers.
However, although Metropolis Monte Carlo methods asymptotically
guarantee finding the thermal equilibrium state, it is unclear whether
an interpretation beyond this has any validity. We view their
procedure as a protein structure optimization approach.  These authors
enumerated all compact (cubic) self-avoiding conformations, which
allow them to determine the global minimum of each random 27mer
sequence generated. They reported \cite{ShFaGuKa91} that three out of
thirty sequences found the known global minimum; energies varied from
-83.7 to -74.6 (in units of $k_BT$).  Using the SGA, we found that
three out of four sequences obtained compact, 100\% cubic
conformations (see fig. \ref{27cubic}), with energies ranging from
-78.1 to -69.9 (units in $k_BT$).  We cannot determine whether they
are the global energy-minimum structures without carrying out a full
enumeration; this is computationally time consuming due to the
NP-complete nature of the problem, and, more importantly, unnecessary
for our design purposes.  Furthermore, we were unsuccessful in our
correspondence with the authors and were unable to obtain the random
interaction matrices specifically used in their work (\cite{ShFaGuKa91}).

\begin{figure*}[!htb]
\centerline{
\psfig{file=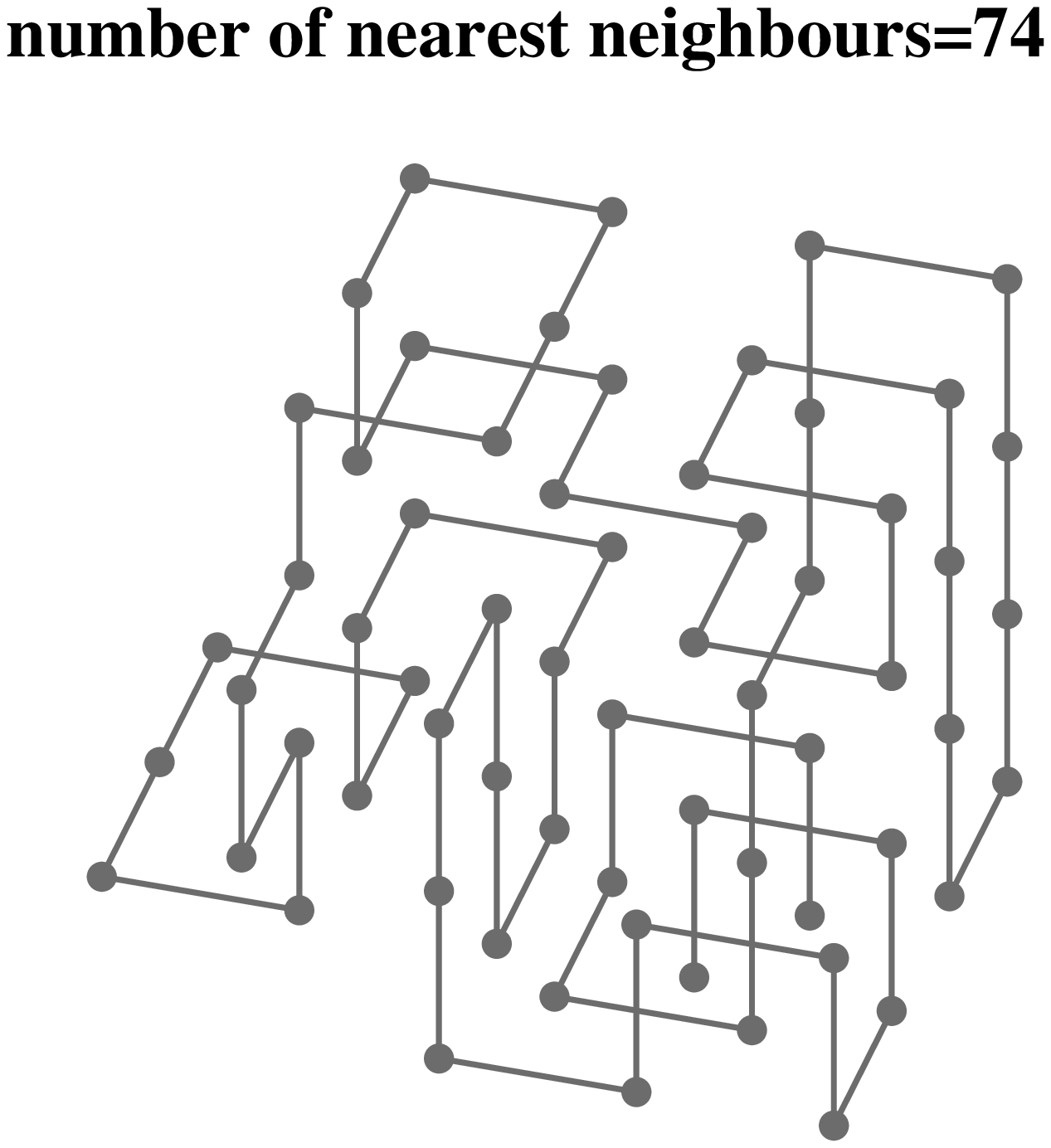, width=3.5in}
\psfig{file=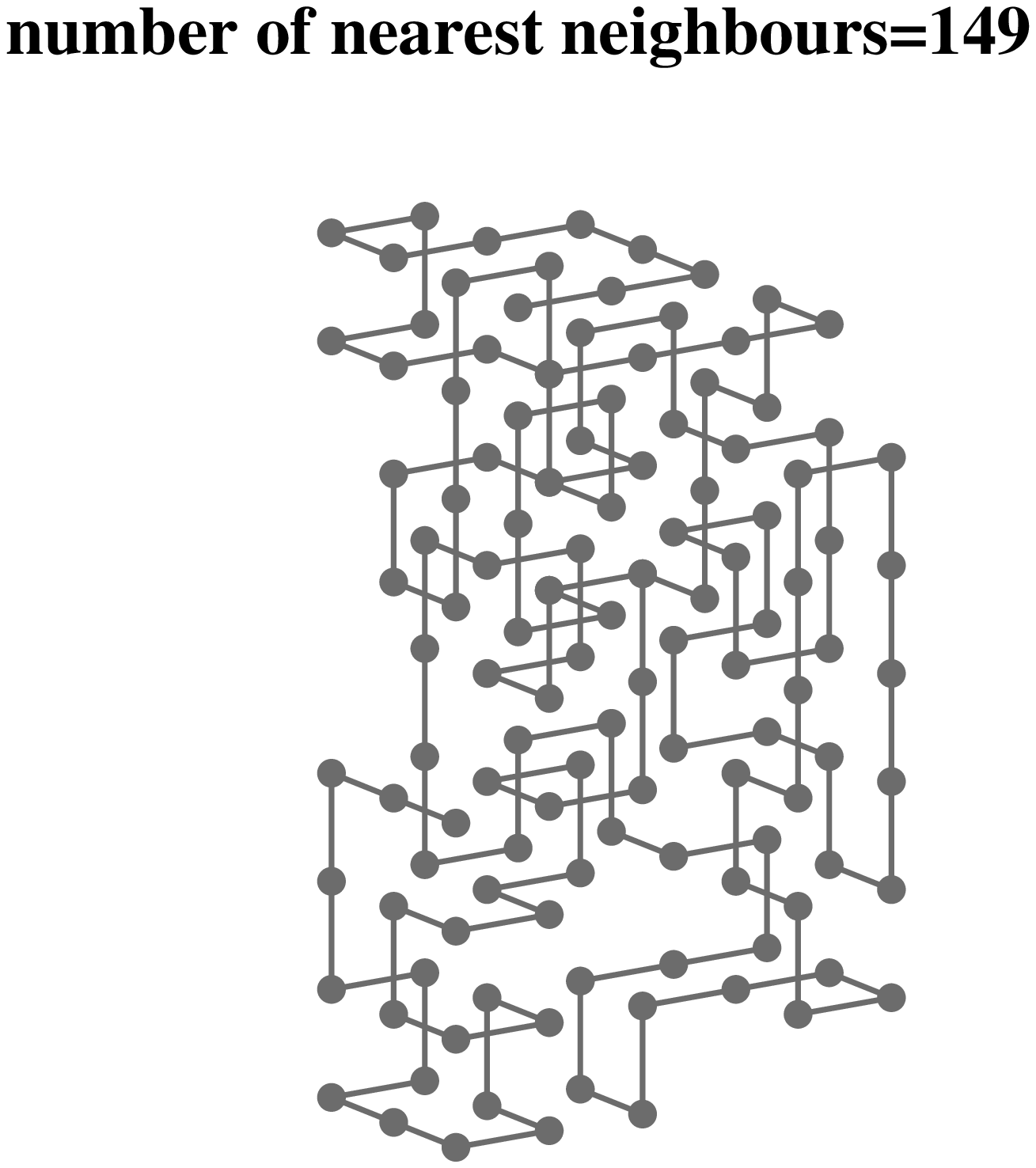, width=3.5in}
           }
\caption{Left: 91\% cubic conformation for a 64mer homopolymer. 
Right: 85\% cubic conformation for a 125mer homopolymer.}
\label{homopolymer}
\end{figure*}

Longer polymers (64mer, 125mer) are more challenging since the
conformational space grows exponentially with the polymer length.  We
continued to use the REM with $B$=0. This corresponds to uniform
interactions and thus guarantees that cubic conformations
(4$\times$4$\times$4, 5$\times$5$\times$5 respectively) occupy the
multiply-degenerate ground states.  The 64mer reached 91\% cubicity and the
125mer runs found a conformation with 85\% cubicity (fig.
\ref{homopolymer}).  Since we are restricting our discussion to cubic
lattices, it is more accurate to describe structures according to how
`cubic' they are, rather than using the more general term
`compactness'.

It is unfortunate that 100\% cubic conformations were not found;
however, we are using the simplest genetic algorithm. Nevertheless,
this initial exercise was useful to establish optimal values for the
GA parameters.  The optimal GA parameters for short polymers were: a
minimum population size of 400; a 20\% probability of crossover and a
4\% probability of mutating a bond direction.  Longer polymers
required a larger minimum population size of 1000; a 90\% probability
of crossover and a 2\% probability of mutation.  Having established a
good set of GA parameters, the SGA was analysed using the second test
function, the HP-model.

Unger and Moult studied random HP-sequences of length 27 and 64
monomers \cite{UngMou93:3d}.  They used two optimization methods: a
Metropolis Monte Carlo (MC) method and a variation of the Monte Carlo
method that incorporates a genetic algorithm (GAMC).  The GAMC
method corresponds to a population of Metropolis Monte Carlo
conformers which `mix' between themselves through a crossover
operation. The comparisons between these methods and our SGA are shown
in tables \ref{SGAvUngMou27} and \ref{SGAvUngMou64}.

Our SGA beat Monte Carlo in 17 of the 20 test sequences and equalled
it in finding low energy conformations for two sequences.  On average,
conformations generated by our SGA were 1.1 energy units lower
than those found by the Monte Carlo method for the
27mers and  8.3 units lower for the 64mers.

\begin{table}[htbp]
\begin{tabular}{lrrrr}
sequence&$E_{SGA}$  &  Num Steps  &  $\Delta E_{MC}$  & $\Delta E_{GAMC}$ \\
\hline
273d.1 & -8         &  2.9E+04    &  -1               & 1 \\
273d.2 & -8         &  2.1E+04    &  1                & 1 \\      
273d.3 & -8         &  1.1E+05    &  -2               & 0 \\
273d.4 & -15        &  2.7E+05    &  -4               & 0 \\      
273d.5 & -7         &  9.6E+03    &  0                & 1 \\
273d.6 & -11        &  9.6E+04    &  -2               & 0 \\    
273d.7 & -11        &  2.2E+04    &  -1               & 1 \\    
273d.8 & -4         &  8.6E+04    &  0                & 0 \\    
273d.9 & -7         &  6.0E+04    & -1                & 0 \\   
273d.10& -10        &  5.6E+04    & -1                & 1 \\    
\hline
\bf AVERAGE & \bf -8.9 & \bf 7.6E+04   &  \bf -1.1    & \bf 0.5
\end{tabular}
\caption{SGA comparisons with Unger and Moult's 27mer results. The
  sequence number corresponds to the number used by Unger and Moult to
  label their HP sequences. $E_{SGA}$ is the
  lowest energy found by the SGA. The `num steps' column reports the
  number of energy evaluations carried out by the SGA to reach the
  lowest energy state. $\Delta E_{MC}$ is the energy difference
  between the lowest energies found by the SGA and Unger \& Moult's
  Monte Carlo procedure: $\Delta E_{MC}=E_{SGA} -E_{MC}$. Similarly,
  $\Delta E_{GAMC}$ is the energy difference between the SGA and Unger
  \& Moult's GAMC method.}
\label{SGAvUngMou27}
\end{table}

\begin{table}[htbp]
\begin{tabular}{lrrrr}
sequence&$E_{SGA}$  &  Num Steps  &  $\Delta E_{MC}$  & $\Delta E_{GAMC}$ \\
\hline
643d.1  &  -21 &  6.4E+05  & -9  & 6 \\
643d.2  &  -26 &  1.5E+06  & -9  & 3 \\
643d.3  &  -36 &  1.7E+06  & -12 & -1 \\
643d.4  &  -30 &  1.9E+06  & -12 & 4 \\
643d.5  &  -28 &  6.3E+05  & -8  & 4 \\
643d.6  &  -22 &  4.6E+05  & -6  & 7 \\
643d.7  &  -17 &  2.1E+05  & -2  & 3 \\
643d.8  &  -28 &  1.4E+06  & -9  & 1 \\
643d.9  &  -29 &  9.8E+05  & -10 & 3 \\
643d.10 &  -20 &  3.1E+05  & -6  & 4 \\
\hline
\bf AVERAGE & \bf -25.7 & \bf 9.8E+05  & \bf -8.3& \bf 3.4
\end{tabular}
\caption{SGA comparisons with Unger and Moult's 64mer results. The
  sequence number corresponds to the number used by Unger and Moult to
  label their HP sequences. $E_{SGA}$ is the lowest energy found by
  the SGA. The `num steps' column reports the number of energy
  evaluations carried out by the SGA to reach the lowest energy state.
  $\Delta E_{MC}$ is the energy difference between the lowest energies
  found by the SGA and Unger \& Moult's Monte Carlo procedure: $\Delta
  E_{MC}=E_{SGA} -E_{MC}$. Similarly, $\Delta E_{GAMC}$ is the energy
  difference between the SGA and Unger \& Moult's GAMC method.}
\label{SGAvUngMou64}
\end{table}

When compared to the GAMC method, the SGA performed on average as well
as the GAMC for the 27mers; SGA conformational energies were on
average 0.5 unit higher.  In the 64mer case, SGA conformations were on
average 3.4 energy units higher than the conformational energies found
by the GAMC.  There was one 64mer sequence for which the SGA found a
lower energy conformation than the GAMC. An important note is the
speed at which the SGA found low energy conformations.  Only a
fraction of the total number of steps were required for the SGA when
compared to the Monte Carlo and GA-Monte Carlo. For example, for the
27mer, to reach a solution of comparable quality, the number of steps
required by the SGA was 3\% of the number of steps required by the MC
method, and 4\% for the GAMC.

\begin{figure}[!htb]
\centerline{
\psfig{file=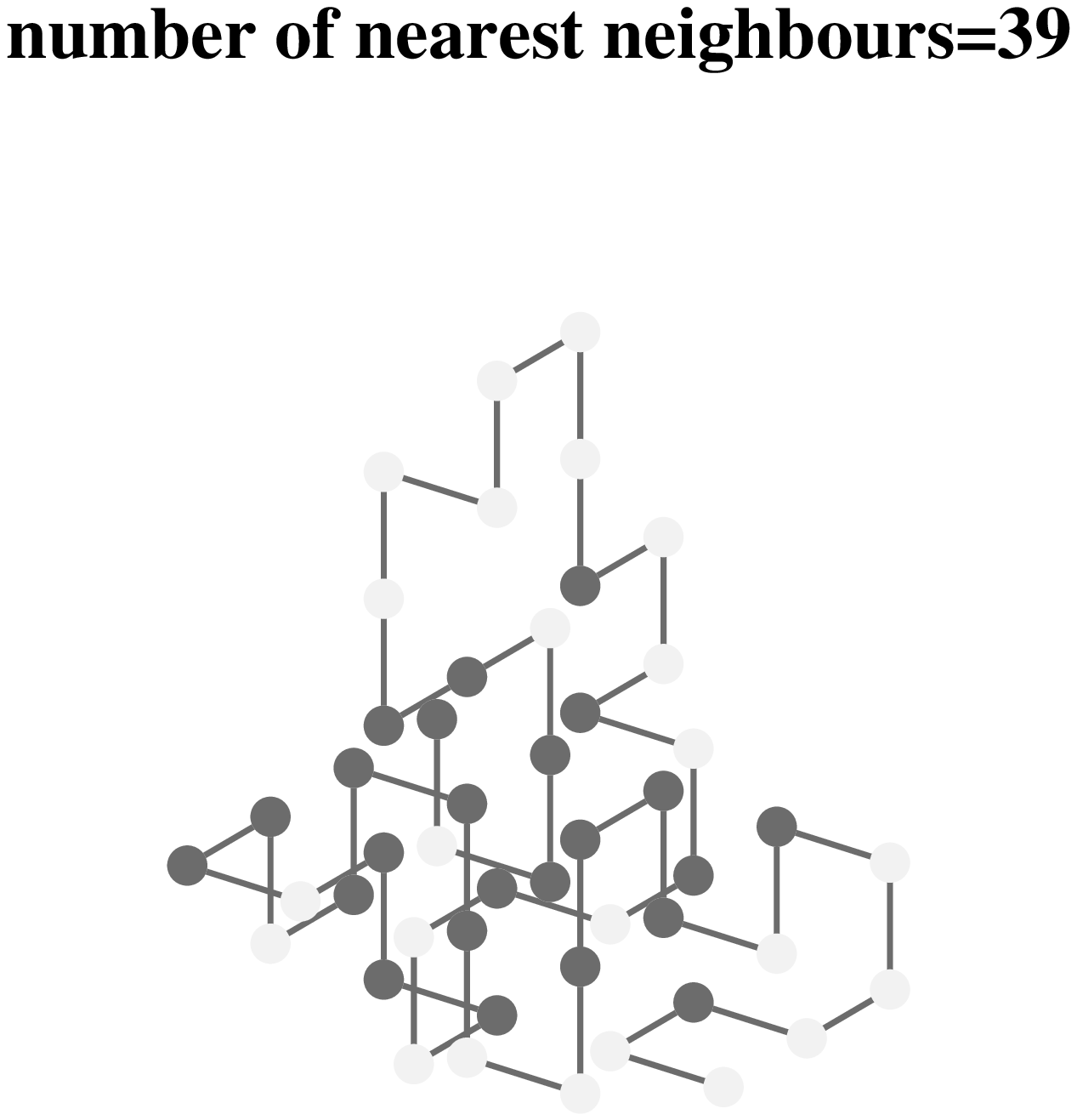, width=3in}
           }
\caption{Example of a compact HP conformation, with a
hydrophobic core, found by the SGA.}
\label{compactHP}
\end{figure}

Further studies were carried out using HP-sequences taken from Yue
\etal \cite{Yueetal95}.  They designed ten 48mer sequences and
determined the native conformational states using a constraint-based
hydrophobic core construction method.  This method determines the
global minima of HP-sequences by constructing conformations with a
core of H (hydrophobic) residues that also minimize the surface area
of the conformation.  The difference in the conformational energies
found by the SGA and the native state energy is labelled as $\Delta
E_N$.  Yue {\etal} also used a conformational search algorithm for
HP-sequences called a `hydrophobic zipper' (HZ).  In this case, the H
monomers attract nearby H monomers and bring them together in a
process akin to nucleation.  The energy difference between
conformations found by this method and the SGA is denoted as $\Delta
E_{HZ}$.  The comparisons are summarised in table \ref{SGAvYueDil48}.
Conformations found by the SGA were on average 6.0 energy units
higher.  An example of a compact conformation by the SGA is shown in
figure \ref{compactHP}. In no cases did the SGA equal or beat
conformational energies obtained by the hydrophobic zipper method.

Runs took from several minutes for short polymers (27mers, 48mers)
to hours for longer cases (64mers, 125mers) on a DEC 3000 Alpha
workstation.

\begin{table}[!htb]
\begin{tabular}{lrrrr}
sequence&$E_{SGA}$  &  $\#$Steps  &  $\Delta E_{HZ}$  & $\Delta E_{N}$ \\
\hline
483d.81  &  -24 &  1.6E+06  & 7  & 8 \\
483d.82  &  -24 &  4.7E+05  & 8  & 10 \\
483d.83  &  -23 &  1.9E+05  & 7  & 10 \\
483d.84  &  -24 &  8.6E+05  & 7  & 10 \\
483d.85  &  -28 &  2.4E+05  & 2  & 4 \\
483d.86  &  -25 &  5.5E+05  & 4  & 7 \\
483d.87  &  -27 &  3.8E+05  & 2  & 5 \\
483d.88  &  -26 &  4.2E+05  & 3  & 5 \\
483d.89  &  -27 &  1.3E+05  & 4  & 7 \\
483d.90  &  -26 &  7.0E+04  & 3  & 7 \\
\hline
\bf AVERAGE & \bf -25.4 & \bf 4.9E+05  & \bf 9.4& \bf 7.3
\end{tabular}
\caption{SGA comparisons with Yue and Dill's 48mer studies.
  Sequence number corresponds to the number used by Yue \etal to label
  their HP sequences. $E_{SGA}$ is the lowest
  energy found by the SGA. The `num steps' column reports the number
  of energy evaluations carried out by the SGA to reach the lowest
  energy state. $\Delta E_{HZ}$ is the energy difference between the
  lowest energies found by the SGA and Yue \etal's hydrophobic zipper
  method: $ \Delta E_{HZ}=E_{SGA} -E_{HZ}$.  Similarly, $\Delta
  E_{N}$ is the energy difference between the SGA and the native state
  energy found by Yue \etal using the constraint-based hydrophobic
  core construction method.}
\label{SGAvYueDil48}
\end{table}

\section{Discussion}
\label{Discussion}
The conformations found by the SGA are not promising in
themselves.  However, the aim of the study is to analyse the simple
genetic algorithm and determine what factors are important in
designing PSP-GAs. What can we learn from these results about PSP-GAs? 

It is clear from the 27mer studies that the SGA performs well for short
polymers and tends to find compact, low energy structures.  However,
as was evident in our initial investigation of 64mers and 125mers
using the REM, the SGA performs below average.  Although the SGA
method was better than the Monte Carlo method, it did not perform as
well as the GAMC and was even worse than the HZ algorithm.  To some
extent, it is not surprising that the hydrophobic zipper method
performed better than the SGA.  HZ is a specialised search algorithm
for the HP-model and generates conformations by explicitly forming H-H
contacts. It is not uncommon for specialised algorithms to solve
certain cases of an NP-complete problem - however, a {\em general}
solution remains difficult \cite{TSP}.

\begin{figure*}[!htb]
\begin{center}
\begin{tabular}{lr}
\psfig{file=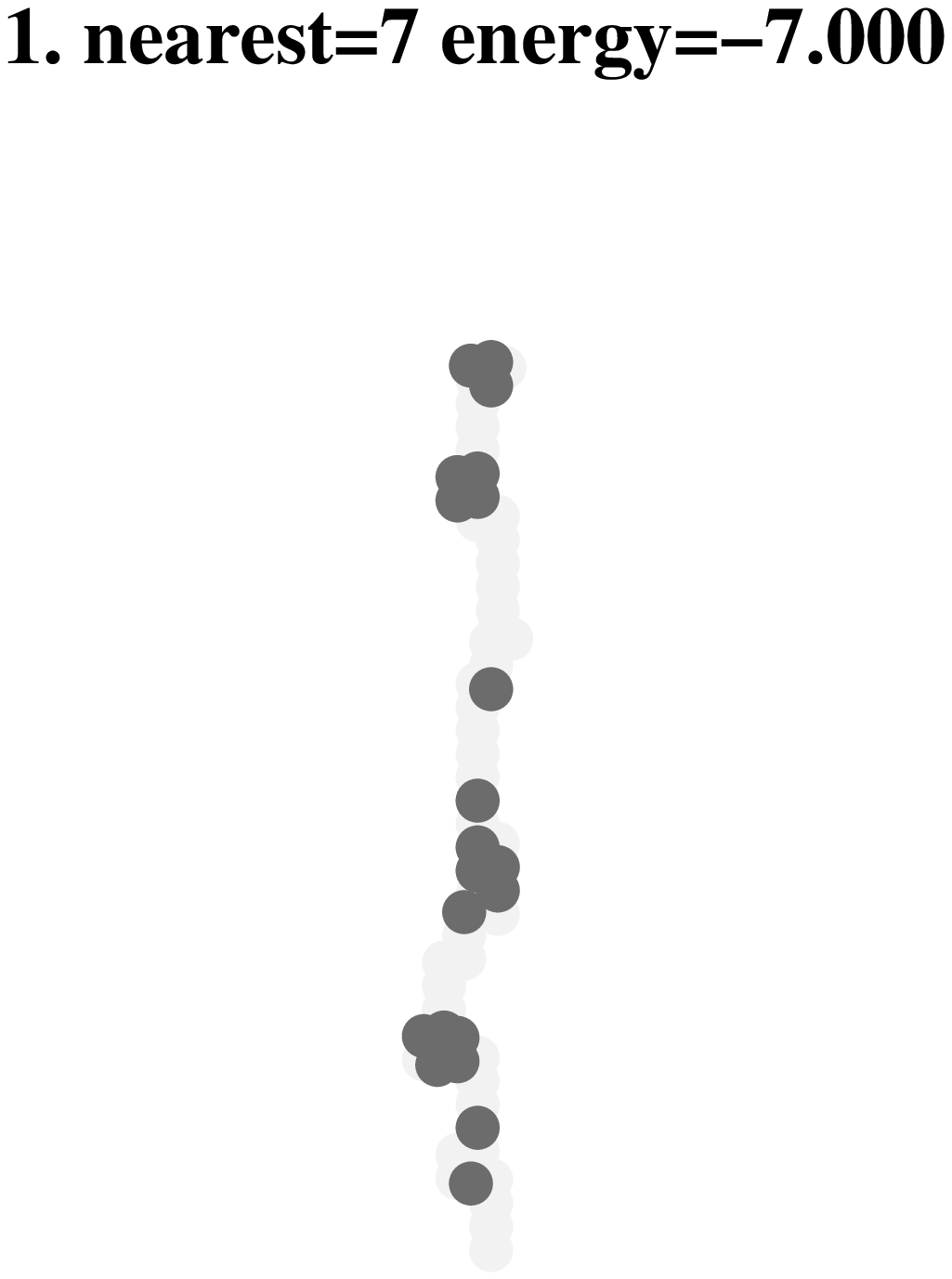,width=3.2in} & 
\psfig{file=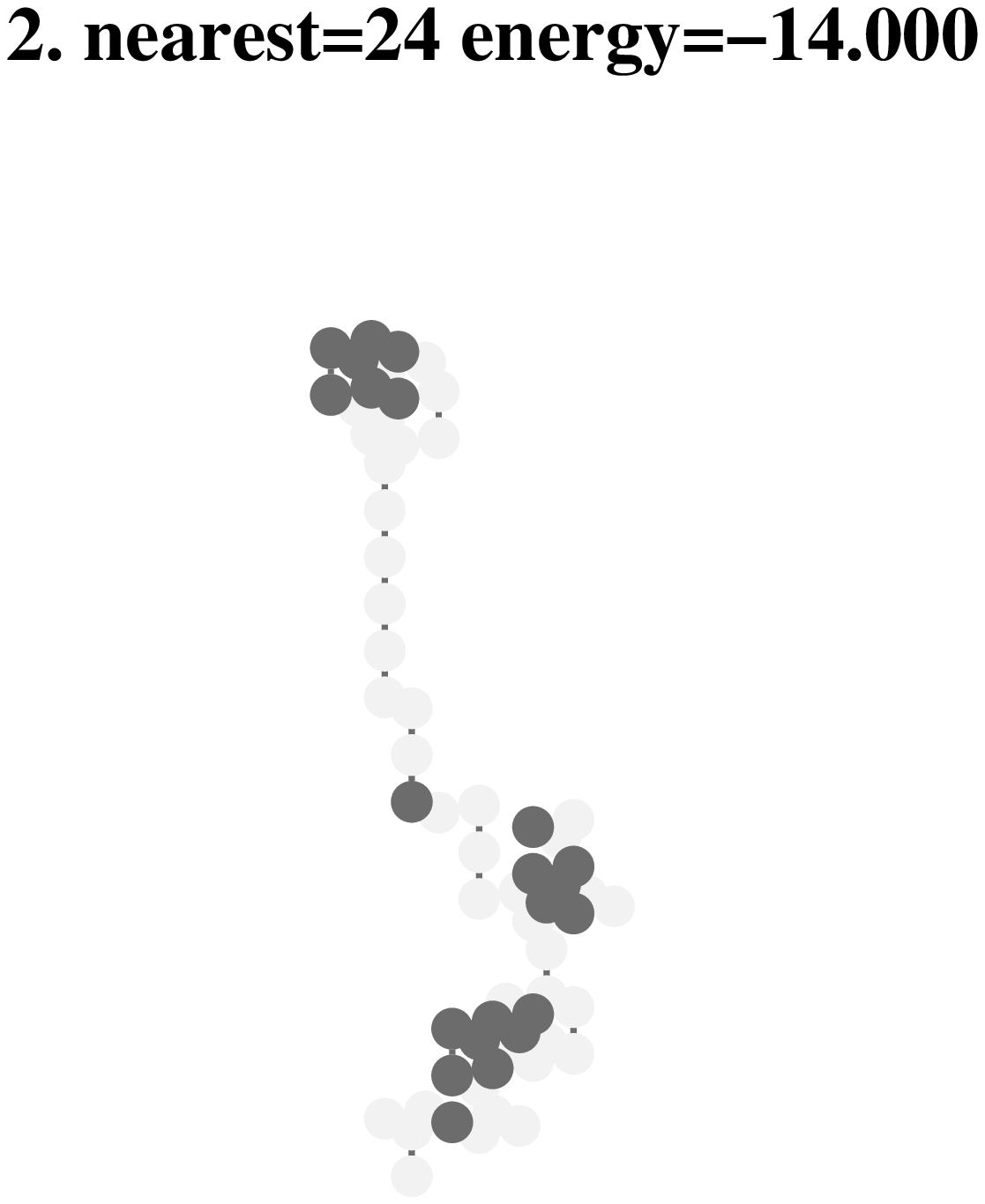,width=3.2in} \\
\psfig{file=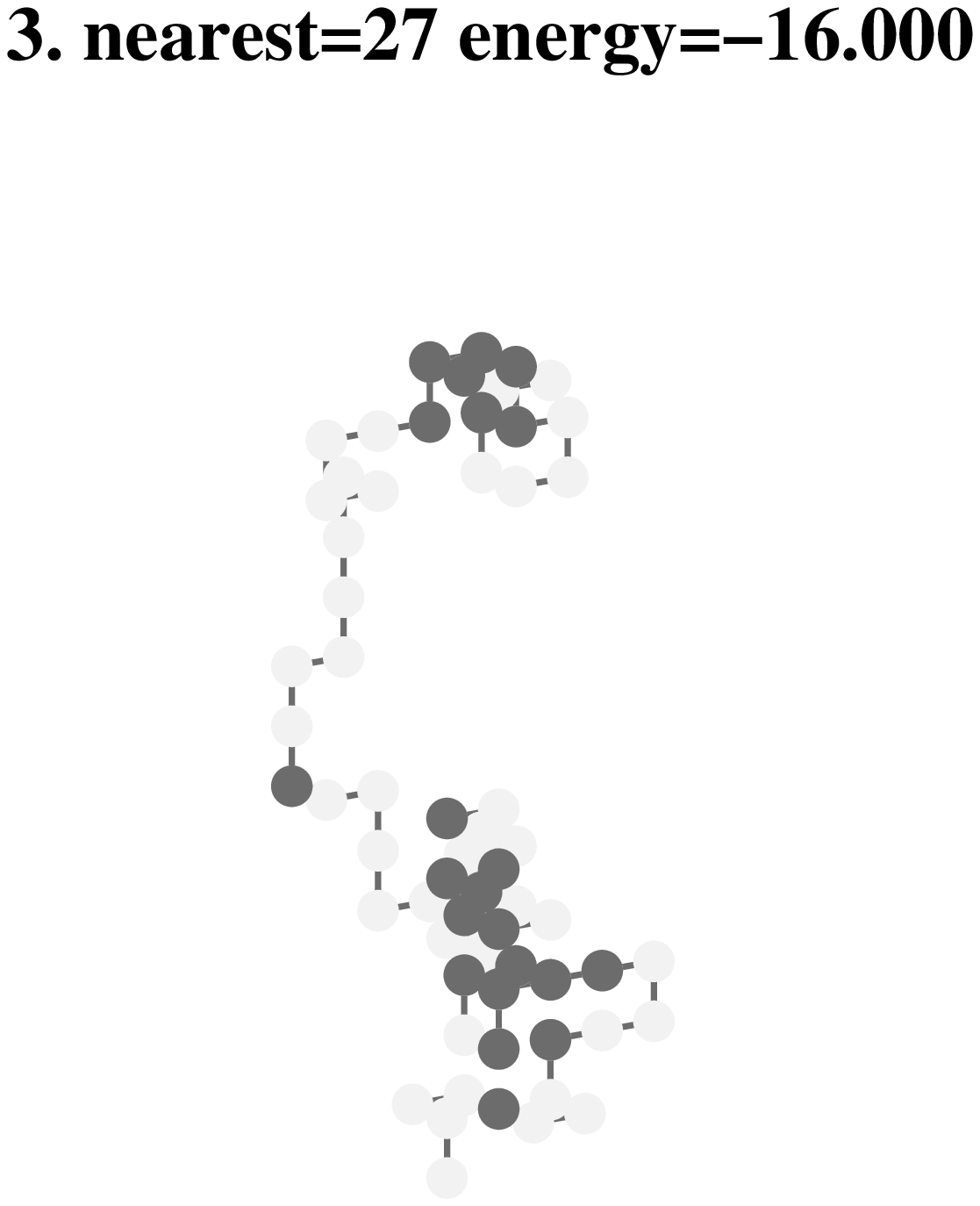,width=3.2in} & 
\psfig{file=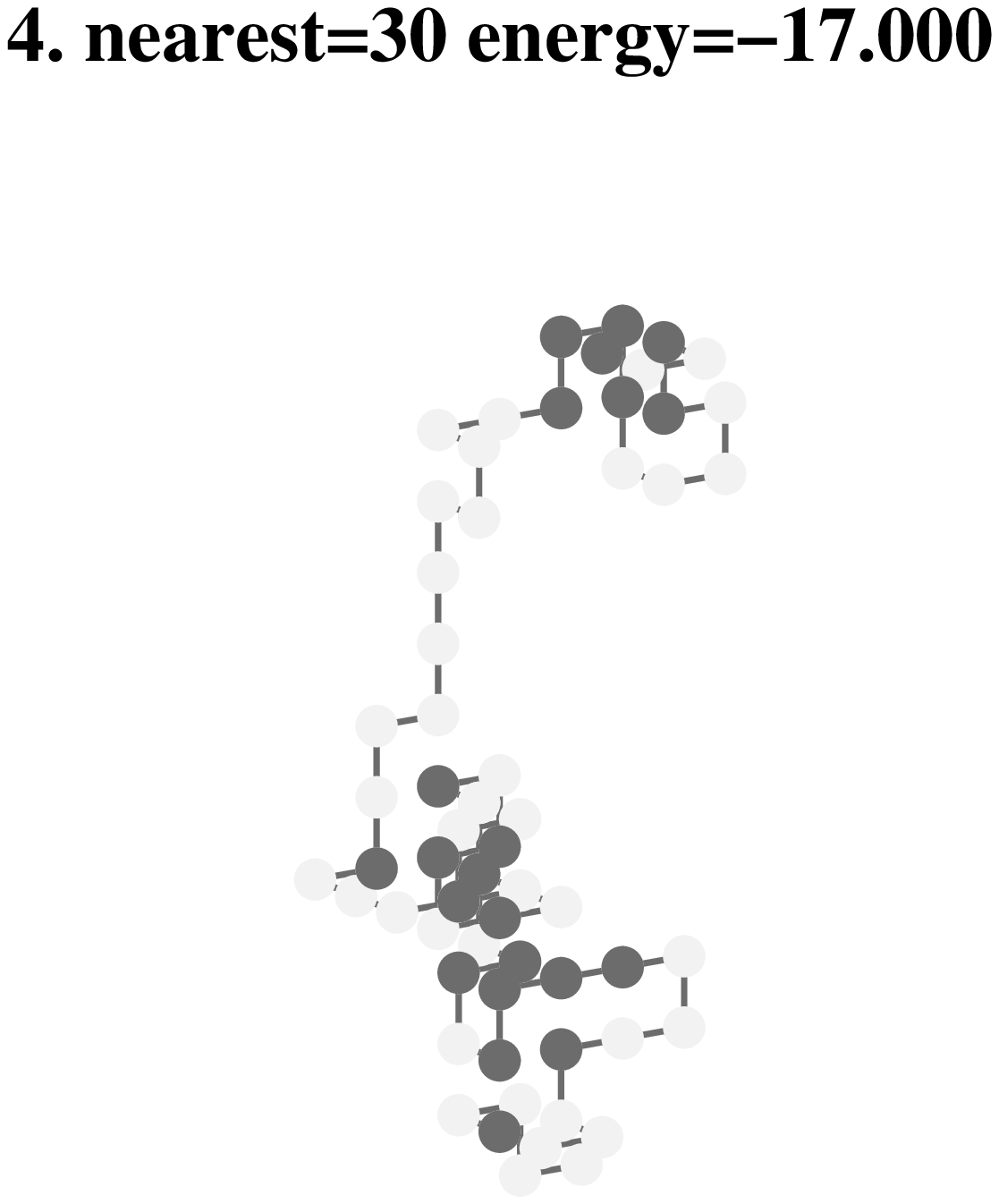,width=3.2in}
\end{tabular}
\end{center}
\caption{Evolution of the best conformation per generation: formation
  of clusters in the Simple Genetic Algorithm. Dark=H=hydrophobic,
  Light=P=polar.}
\label{cluster}
\end{figure*}

In general the SGA finds it hard to generate compact conformations for
the longer polymers - why? We investigated this further using the
HP-model. The best way to analyse the dynamics of the SGA is to
observe its time evolution rather than merely the end result.  Thus,
the lowest energy conformations were observed at various time points.
In several cases, the SGA produced two clusters of residues with
hydrophobic (H) cores connected by a `thread' - similar to loop
regions in real proteins.  It is promising that such conformations
(see figure \ref{cluster}) are possible since proteins with
sub-structures connected by loop regions are common in Nature.
However, this is not the lowest energy conformation for the HP-model,
so we conclude that the clustering is due to restrictions imposed by
the GA dynamics, that is to selection and recombination.  Selection is
an important aspect of the GA dynamics but we do not think it is the
reason for the occurrence of clustered solutions; if a lower energy
globule existed in the population, the selection function would have
assigned it a high probability and elitism would guarantee its
selection into the next generation.  

The problem must therefore lie in the recombination stage: it appears
that the one-point crossover and mutation operations cannot
inter-digitate the clusters successfully.  We believe that these
recombination operators would fold the clusters {\em into} each other
creating many sites with multiple occupancies; these are penalised by
the excluded volume terms in Eq.\ref{HP} heavily reducing the
`survival probability' of such structures.  The difficulty appears to
be in the formation of only two large clusters, or `building blocks';
we think that more clusters but smaller in size would be easier to
manipulate. How do we promote this in the GA dynamics?

Genetic algorithms manipulate partial solutions in their search for
the overall optimal solution \cite{Hol75,Gol89:gasoml}.  These partial
solutions or `building blocks' correspond to sub-strings of a trial
solution - in our case local sub-structures within the overall
conformation.  Clearly, the level of crossover influences the size of
the building blocks.  A multi-point crossover would generate smaller
building blocks (a higher `resolution') and consequently smaller
clusters that should facilitate their inter-digitation. In the case of
the 27mer, the one-point crossover generates building blocks of 13.5
monomers, on average.  Based on this argument, 48mers and 64mers
require at least two to three point crossovers to reduce the size of
the building blocks to $\sim10$.  Furthermore, the local perturbation
dynamics of the Monte Carlo methods aids the local `fine tuning' of
conformations as manifested by the GAMC method.  It is believed that
the two recombination operators of multi-point crossover and a local
perturbation are required for any PSP-GA to be fully effective.  Work
is under way to design a PSP-GA to optimize real protein structures
(rather than lattice models) which includes these requirements.

One further problem with genetic algorithms is `knowing when to stop'.
Most optimization algorithms deal with a single solution at a time and
decide to stop when there has been no change in the cost or energy
function for a successive number of steps.  Since GAs deal with an
ensemble of solutions, a quantity analogous to the statistical
mechanical free energy is used. The `population free energy', $F$, is
calculated from its `partition function', $Z$:

\begin{equation}
Z = \sum_{i=1}^{N}{e^{-E_i}}
\end{equation}

\noindent where the sum is over the total number of conformers in a population
and $E_i$ is the energy of the $i^{th}$ conformer. Hence,

\begin{equation}
F = - \ln (Z)
\end{equation}

\begin{figure}[!htb]
\centerline{
\psfig{file=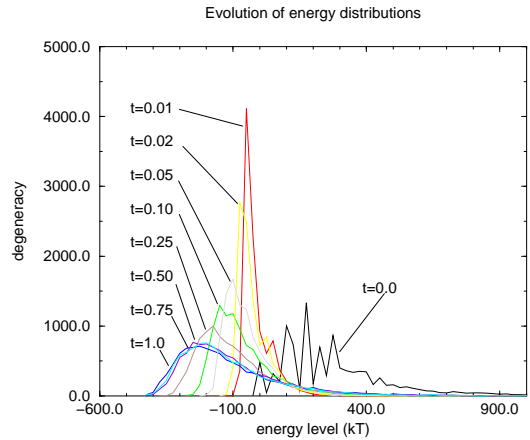, width=3in}
           }
\caption{Evolution of a population: plotting the energy distributions
  at various time steps. $t$ is the percentage of the total
  run-time elapsed. The population converges within 50\% of the run time.}
\label{distrib}
\end{figure}

\begin{figure}[!htb]
\centerline{
\psfig{file=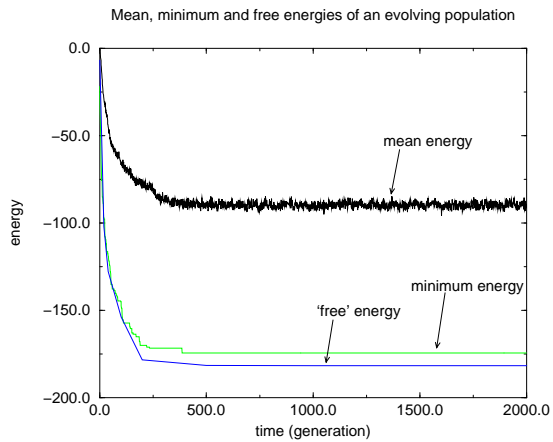, width=3in}
           }
\caption{Evolution of the energies of a population. The free energy,
  which characterises the energy distribution of a population, reaches
  a convergence, or equilibrium point.}
\label{energies}
\end{figure}

This approach is advantageous over using the mean energy of the
population in two ways.  Firstly, the mean energy fluctuates around
the equilibrium making it difficult to use as a stop criterion, and
secondly, $F$ contains more information on the shape of the energy
distribution, including, in particular, its `entropy'. $F$ appears to
play the role of a Lyapounov function in the GA dynamics - convergence
is synonymous with its minimal value.  Figure \ref{distrib} plots the
energy distributions of the population at various time steps.  The
plot shows very clearly that the GA dynamics converge the population
of conformations to an equilibrium distribution. This is characterised
by $F$ as shown in figure \ref{energies}.

In conclusion, the genetic algorithm approach to the protein structure
prediction problem offers a promising potential method of solution.
GAs are fast and efficient at searching the rugged conformational
landscapes presented by protein molecules.  We have established some
guidelines for designing PSP-GAs in this paper and are currently
implementing them in an improved GA to search for realistic protein
structures.

\section{Acknowledgements}
MK would like to thank Paul van der Schoot and Steven Brenner for
interesting discussions. The authors would also like to thank Ken Dill
and Kaizhi Yue for various information. MK is financially supported by
an EPSRC award.


\bibliographystyle{plain}

\end{document}